\documentclass[proof]{WileyASNA-v1}

\usepackage{graphicx}
\articletype{Article Type}%

\received{}
\revised{}
\accepted{}

\def\loggf{$\log\,gf$}
\def\Teff{$T_{\rm eff}$}
\def\logg{$\log\,g$}

\def\Vt{$V{\rm t}$}

\def\kms{km s$^{-1}$}

\begin{document}

\title{NLTE CNO abundances in a sample of nine field RR lyr type stars}

\author[1]{Sergei M. Andrievsky}

\author[2,3]{Sergey A. Korotin}

\author[1]{Valery V. Kovtyukh}

\author[4]{Sergey V. Khrapaty}

\author[5]{Yuri Rudyak}

\authormark{ANDRIEVSKY \textsc{et al}}

\address[1]{\orgdiv{Astronomical Observatory}, \orgname{Odessa National University, Ministry of Education and Science}, \orgaddress{\state{Shevchenko Park, 65014, Odessa}, \country{Ukraine}}}

\address[2]{\orgdiv{Stellar Physics Department}, \orgname{Crimean Astropnysical Observatory}, \orgaddress{\state{298409, Nauchny}, \country{Republic of Crimea}}}

\address[3]{\orgdiv{Department of stellar spectroscopy and nonstationary stars}, \orgname{Institute of Astronomy, Russian Academy of Sciences}, \orgaddress{\state{RU-119017, Moscow}, \country{Russia}}}

\address[4]{\orgdiv{Department of Computer Systems and Technologies}, \orgname{Interregional Academy of Personnel Management}, \orgaddress{\state{Frometivska str., 2, 03039, Kyiv}, \country{Ukraine}}}

\address[5]{\orgdiv{Department of Medical Physics of Diagnostic and Therapeutic Equipment}, \orgname{I. Horbachevsky Ternopil National Medical University}, \orgaddress{\state{Maidan Voli, 1, 46001, Ternopil}, \country{Ukraine}}}

\corres{S.M. Andrievsky, 
Astronomical Observatory, Odessa National University, Shevchenko Park, 65014, 
Odessa, Ukraine \email{andrievskii@ukr.net}}

\abstract{

For the first time, a direct NLTE analysis of carbon and nitrogen lines in the spectra of 
nine RR Lyrae stars was carried out. We have determined the abundances of these elements 
together with oxygen, and have shown that the nitrogen content is increased in metallicity
deficient program stars. We conclude that this is a sign of the first dredge up, which 
occurred at the previous stage of the red giant branch, and brought material processed 
in an incomplete CNO cycle to the surface of the star. This effect is significantly 
enhanced by thermohaline (extra-) mixing, which is more effective for metal-poor RR Lyrae stars.
This is clearly seen in the plot showing that C/N ration in our sample of stars gradually
decreases as metallicity decreases from about --0.2 to --2. Oxygen abundance depends on metallicity 
in a similar way to what we see in the Population II stars.

}


\maketitle

\section{Introduction}

RR Lyrae stars are pulsating giants of A--F spectral classes located
on the Horizontal Branch (HB), which cover the rather wide range of
metallicities from nearly solar to about [Fe/H]= --3 dex (the most
metal-poor field RR Lyr star that we investigated in our series of
papers is UY Boo; it has metallicity --2.73, see \citealt{Wall2009}).

RR Lyrae stars can be found in globular clusters, as well as in the
Galactic  field (halo, thick disc and bulge).
They are low-mass stars (less that one solar mass), the evolutionary
stage of which is determined by the core
helium burning, when the helium nuclei are fusing to carbon and oxygen
nuclei. In addition, protons are fusing to 
$\alpha$-particles in a shell surrounding the stellar core.
At the next stage of evolution, the HB star with a carbon core
passes to the region of the lower temperatures in the HR diagram.
As an asymptotic giant branch star, it has two shell sources of
energy release -- helium and hydrogen burning shells.

RR Lyrae stars invariably attract attention of astrophysicists engaged
in photometric and spectroscopic studies.   
As a rule, the abundances of $\alpha$-elements, iron-peak elements and
sometimes heavy elements, like strontium, yttrium, barium, were derived
in these stars (see, e.g. recent papers by \citealt{Magurno2018,Magurno2019}).

Of  particular interest is the determination in RR Lyrae stars of the
abundances of such chemical elements as
carbon, nitrogen and oxygen. For instance, the LTE abundanhces of C and N
were derived by \cite{Butler1982}.
These authors showed that the [C/Fe] ratio for 19 field RR Lyrae stars are
typical for  unevolved Population II stars. Nitrogen abundance (with
indicated uncertainty) was determined only for three stars, and only
the upper limit was estimated for the rest of the stars. Later \cite{Butler1986}
reported about NLTE oxygen abundance determination from the near
IR O~{\sc i} triplet for the same sample of stars. It should be noted
that the reported dependence of [O/Fe] on [Fe/H] (see their Fig. 1)
looks rather strange: as [Fe/H] decreases, [O/Fe] decreases too, which is
opposite to what is expected for the halo Population II stars.

According to  \citet{Smith1995} the overwhelming majority of the field RR Lyrae 
stars are not components of the binary systems, but recent study of
\citet{Kervella2019} based on Gaia DR2 shows that at least 7\% of RR Lyrae
stars are binaries.  
 
Carbon enhanced metal-poor stars were discovered and studied among RR Lyrae
objects (e.g. \citealt{Kinman2012}, \citealt{Kennedy2014}, \citealt{Reggiani2014},
\citealt{Xia2019,Xia2020}).
Those stars are believed to gain high carbon abundance from either
thermally pulsing AGB stars, especially from initially more massive
companion star in a binary system due to the contamination of  their surface
with processed material. For example, SDSS J170733.93+585059.7 has
been identified as the RR Lyr star in the binary system with an
extremely high relative carbon abundance [C/Fe] = +2.79
\citep{Kinman2012}.

\cite{Takeda2006} performed NLTE analysis in order to derive oxygen
abundance in a sample of the four RR Lyrae 
stars. The authors obtained an increased oxygen abundance in their
program stars, [O/Fe] is in the range from about 0 to +1, as expected
for metal-poor objects. A similar result was obtained by
\citet{Andrievsky2018}, who derived NLTE oxygen abundances
in a sample of 30 Galactic field RR Lyrae stars.
[O/Fe] were distributed in the range from 0 to about 1 dex with a typical
for the Population II stars dependence on [Fe/H] (metalliciy
is in the range from 0 to about --3 dex).

In our recent paper \citep{Andrievsky2020} we reported on the results
of the determination of C and O NLTE abundances in a sample of 21 RR Lyrae
stars from the Galactic field. In all  stars studied, carbon showed
underabundance comparing  to the solar (C/H), while oxygen was found
to be overabundant. Its relative abundance [O/Fe] shows a clear
dependence on [Fe/H], as expected for the Galactic halo stars. 

The theory of stellar evolution predicts that the processes of internal
nucleosynthesis in HB stars can
only change the surface abundances of such light elements as carbon, nitrogen
and oxygen. Since nitrogen
abundance in RR Lyrae stars was practically not studied in the past, we decided
to carry out the comprehensive
NLTE analysis of the carbon, nitrogen and oxygen lines in the spectra of 9
Galactic field RR Lyrae stars.

\section{Observations}

With the Apache Point Observatory (USA, APO) 3.5-m telescope we observed nine
RR Lyrae stars.  All spectra have a resolution of 31,000, ranging from from 3,900 to
10,400 \AA, with S/N from 70 to 200. Spectra of rapidly rotating
hot stars were obtained to facilitate the elimination of atmospheric
absorption lines (except very strong lines). Table \ref{obs} shows a list of
target stars observed at APO and their properties.

\begin{table*}
\center
\caption[]{Characteristics of the program stars and abundances
of iron, carbon, nitrogen and oxygen. For the reference we give here our NLTE
absolute abundances (X/H) of carbon, nitrogen and oxygen in the Sun: 8.43, 7.89 and 8.71 respectively.}
\begin{tabular}{cccccccrrrr}
\hline
Star    & Period &  Type &   JD    &\Teff &\logg &  \Vt  & [Fe/H] & [C/Fe]& [N/Fe]& [O/Fe] \\
        &  (d)   &     & 2\,450\,000+& (K) &$g$ in cm/s$^{2}$&\kms&  &       &       &       \\
\hline
DH Peg  & 0.2555 & RRc  & 5192.599 & 6660 & 2.0  &   1.8 & --1.40 & --0.16& +1.21 & +0.68 \\
RU Psc  & 0.3904 & RRc  & 5192.736 & 6420 & 2.0  &   1.8 & --2.03 & --0.41& +1.06 & +0.78 \\
AV Peg  & 0.3904 & RRab & 4995.881 & 6390 & 2.5  &   2.3 & --0.15 &  +0.07& +0.38 & +0.38 \\
RR Gem  & 0.3973 & RRab & 5192.782 & 6750 & 2.3  &   2.3 & --0.22 &  +0.08& +0.30 & +0.51 \\
KX Lyr  & 0.4409 & RRab & 5280.975 & 6380 & 2.7  &   2.3 & --0.24 & --0.23& +0.38 & +0.20 \\
RR Leo  & 0.4524 & RRab & 4905.848 & 6400 & 2.0  &   2.0 & --1.38 & --0.01& +0.73 & +0.82 \\
DX Del  & 0.4726 & RRab & 5192.571 & 6590 & 2.5  &   2.3 & --0.12 & --0.09& +0.00 & +0.30 \\
RR Cet  & 0.5530 & RRab & 5047.397 & 6790 & 2.1  &   2.6 & --1.48 &  +0.04& +0.73 & +0.97 \\
RX Eri  & 0.5872 & RRab & 5521.789 & 6180 & 2.6  &   2.2 & --1.08 & --0.06& +0.89 & +0.91 \\
\hline 
\end{tabular}
\label{obs}
\end{table*}

The prelimenary data reduction was accomplished using programs in the
IRAF \footnote{IRAF is distributed by the National Optical Astronomy
Observatories, which are operated by the Association of Universities
for Research in Astronomy, Inc., under cooperative agreement with the
National Science Foundation} package.

\section{Abundance Analysis}

\begin{figure*}[t]
\centering
\includegraphics[width=1.7\columnwidth]{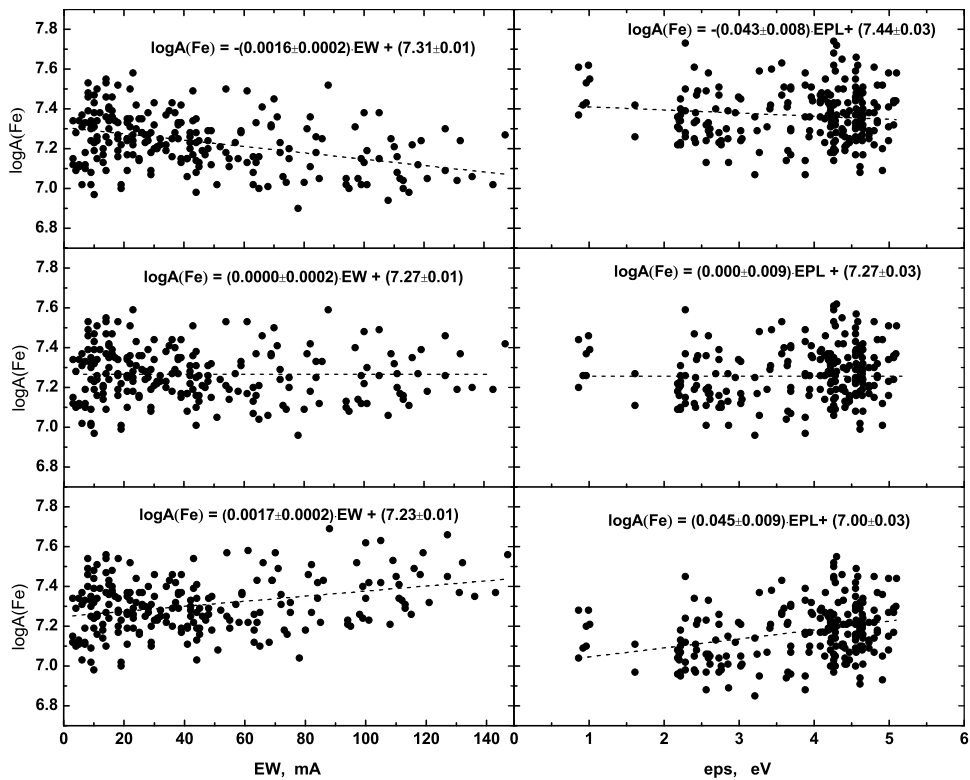}
\caption{This explains the procedute of \Teff\ and \Vt\ determination for KX Lyr
and error estimation for adopted atmosphere parameters. Left panel: middle --
the best \Teff choice, top -- \Teff + 150 K, bottom -- \Teff -- 150 K. Right panel:
middle -- the best \Vt choice, top -- \Vt + 0.2 km/s, bottom -- \Vt --0.2 km/s.}
\label{KX_Lyr_par}
\end{figure*}

\begin{figure*}[t]
\centering
\includegraphics[width=1.7\columnwidth]{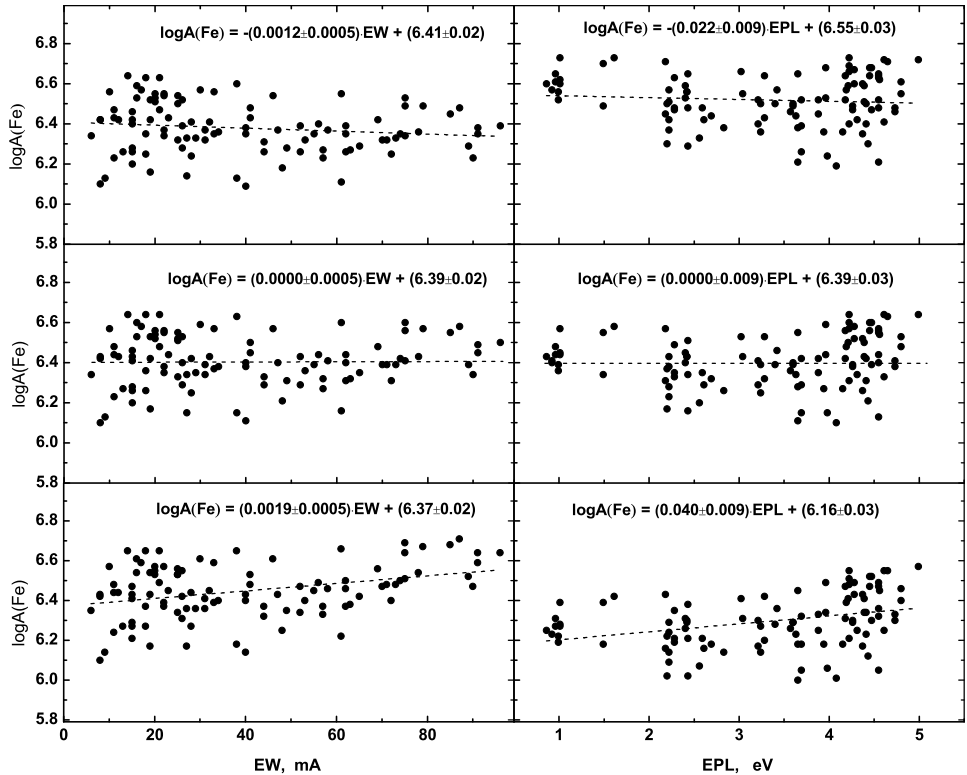}
\caption{The same as Fig. \ref{KX_Lyr_par} but for RX Eri.}
\label{RX_Eri_par}
\end{figure*}

For abundance analysis we employed \citet{Kurucz1993} atmosphere models.
Atmosphere parameters for each program stars were determined by using 
Fe~{\sc i} and Fe~{\sc ii} lines. As it is commonly adopted in spectroscopic 
analysis, the effective temperature was derived by requiring that there be no 
dependence between iron abundance from individual Fe~{\sc i} lines and their 
excitation potentials. Microturbulent velocity was derived by avoiding 
dependence between iron abundance from individual Fe~{\sc i}
lines and their equivalent widths. Finally, the gravity value was found 
by keeping equal mean iron abundances from Fe~{\sc i} and Fe~{\sc ii} 
lines. The corresponding mean values of iron abundance from Fe~{\sc i} 
and Fe~{\sc ii} lines give the final [Fe/H] value  for a given star. Error 
analysis was made in the same way as it was described in \cite{Andrievsky2020}. 
In Fig. \ref{KX_Lyr_par} and Fig. \ref{RX_Eri_par} we graphycally show the 
procedure of \Teff\ and \Vt\ determination for KX Lyr and RX Eri, and sensitivity 
of the adopted \Teff\ and \Vt\ values to variations within $\pm 150$, K and 
$\pm 0.2$ \kms. The change in slopes in both cases can be traced visually.
It should be noted that as a reference point we adopted the solar abundance 
of iron (Fe/H) = 7.48 \citep{Palme2014}.

The total list of the iron lines used, their atomic data and equivalent widths are given 
in the Tables \ref{spectra} and \ref{EWs} in Appendix (the full tables are available on-line).

It should be noted that for most of the stars of our program, the atmosphere 
parameters were first determined in one of our previous work by \citet{Andrievsky2018} 
using the SME code \citep{Piskunov2017}. These parameters were used as a first 
approximation for the current standard abundance analysis. 
For half of the stars, the parameters used turned out to be a good choice, but for 
the rest of the stars, the parameters were refined.

\section{Carbon and Oxygen NLTE abundances}

Atomic models of carbon and oxygen, which were used to derive NLTE abundances 
of these elements in our program stars, are described in details in several papers, see, 
for instance, \cite{Andrievsky2020} for reference. In that paper we also give the lists 
of analyzed carbon and oxygen lines, and a description of how we applyed spectral 
synthesis to match the observed profiles of selected lines of the studied ions.

\section{Nitrogen Abundance Calculations}

To find atomic level populations for the  atom N~{\sc i},
we employed the code MULTI \citep{Carlsson1986}.
For our aim, this code was modified and adapted by
\cite{Korotin1999}. MULTI gives a possibility
to calculate a single line NLTE profile. It should be noted that
the lines of interest as a rule are blended in the real stellar spectra.
In order to take the blending into account, we first calculate with MULTI
the NLTE departure coefficients for those levels that form the line of
interest, and then we include these coefficients in the LTE synthetic
spectrum code SYNTHV \citep{Tsymbal1996}. This enables one to calculate the
source function and opacity for each studied line. Simultaneously, the
blending lines are calculated in LTE with the help of line list and
corresponding atomic data from VALD \citep{Ryabchikova2015}
in the wavelength range of the line under study. This technique is the
same as we used for carbon and oxygen lines \citep{Andrievsky2020}.

For the NLTE calculation we use the model of  N~{\sc i} atom that was
described in \cite{Lyubimkov2011}. That model was
supplemented with an account for the updated atomic data. In particular,
electron collisional rates for the lower 27 levels of N~{\sc i} were taken from
\cite{Wang2014}, who used the detailed quantum mechanics calculations for
this aim. 

The adopted nitrogen model atom consists of 39 N~{\sc i} levels,
49 N~{\sc ii} levels and the ground state of 
N~{\sc iii}. Moreover, additional atomic levels of LTE populations
were included in the equation of the particle number conservation.
Among them there are 66 N~{\sc i} levels and four excited N~{\sc iii} levels.

Nitrogen abundance in the solar atmosphere was found with the help of 
equivalent widths of the eight  N~{\sc i} lines observed at the center of the solar
disc \citep{Grevesse1990, Biemont1990}. The authors report that obtained equivalent 
widths were cleaned of the molecular lines. Our solar nitrogen abundnace (N/H) =
7.89 $\pm$ 0.04 is in good agreement with the result of \citet{Caffau2009}, 
who gives (N/H)=7.86 $\pm$ 0.12, as well as with result of 
\citet{Lodders2019}, who gives (N/H) = 7.85 $\pm$ 0.12.

More information about nitrogen abundance from individual lines in the solar
spectrum can be found in Table \ref{NSolar}.

\begin{table}
\caption[]{Nitrogen abundance in the Sun from different N~{\sc i} lines. NLTE corrections are also given.}
\center
\begin{tabular}{ccrr}
\hline
Line       &EW(m\AA) &	 (N/H) & dNLTE \\
\hline
7442.29 &	2.6         & 7.84	 &--0.02  \\
7468.31 &	4.9         &	7.92	 &--0.02  \\
8216.33 &	8.6         &	7.90	 &--0.03  \\
8242.38 &	3.9         &	7.89	 &--0.03  \\
8629.23 &	4.5         &	7.84	 &--0.03  \\
8683.40 &	7.8         &	7.85	 &--0.04  \\
8718.83 &	4.2         &	7.93	 &--0.03  \\
9392.79 &	9.5         &	7.96	 &--0.03  \\
\hline
\end{tabular}
\label{NSolar}
\end{table}

To derive NLTE abundances in our program stars we used 12 lines of N~{\sc i}
that belong to the four different multiplets. Their parameters were taken
from the VALD \citep{Ryabchikova2015}, and they are given
in Table \ref{Nlines}. All investigated lines to some extent suffer from the
NLTE effects. Generally, observed lines become stronger due to the
NLTE influence. To describe in LTE observed profiles of all 12 lines
with one adopted  nitrogen abundance is practically impossible. 

\begin{table}
\caption[]{N~{\sc i} lines.}
\center
\begin{tabular}{ccrr}
\hline
 $\lambda$ (\AA) & $\chi$ (eV) & \loggf  \\
\hline
 7442.30  &  10.33  & --0.40 \\
 7468.31  &  10.34  & --0.18 \\
 8184.86  &  10.33  & --0.30 \\ 
 8188.01  &  10.33  & --0.29 \\ 
 8216.34  &  10.34  &   0.13 \\ 
 8223.13  &  10.33  & --0.27 \\ 
 8629.24  &  10.69  &   0.08 \\ 
 8683.40  &  10.33  &   0.10 \\ 
 8686.15  &  10.33  & --0.28 \\ 
 8703.25  &  10.33  & --0.31 \\ 
 8711.70  &  10.33  & --0.23 \\ 
 8718.84  &  10.34  & --0.35 \\ 
\hline
\end{tabular}
\label{Nlines}
\end{table}

\begin{figure*}[t]
\centering
\includegraphics[width=0.66\columnwidth]{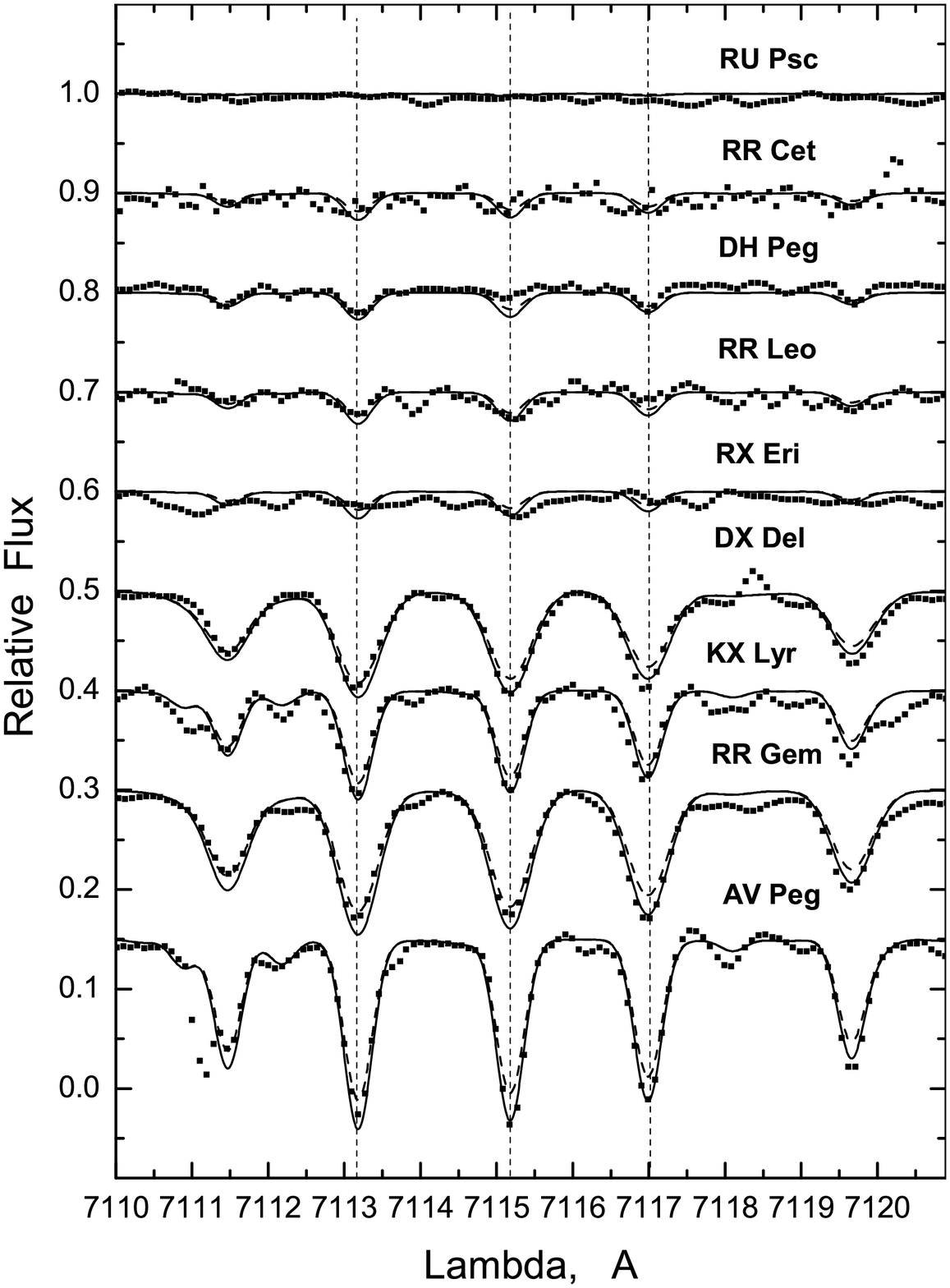}\includegraphics[width=0.66\columnwidth]{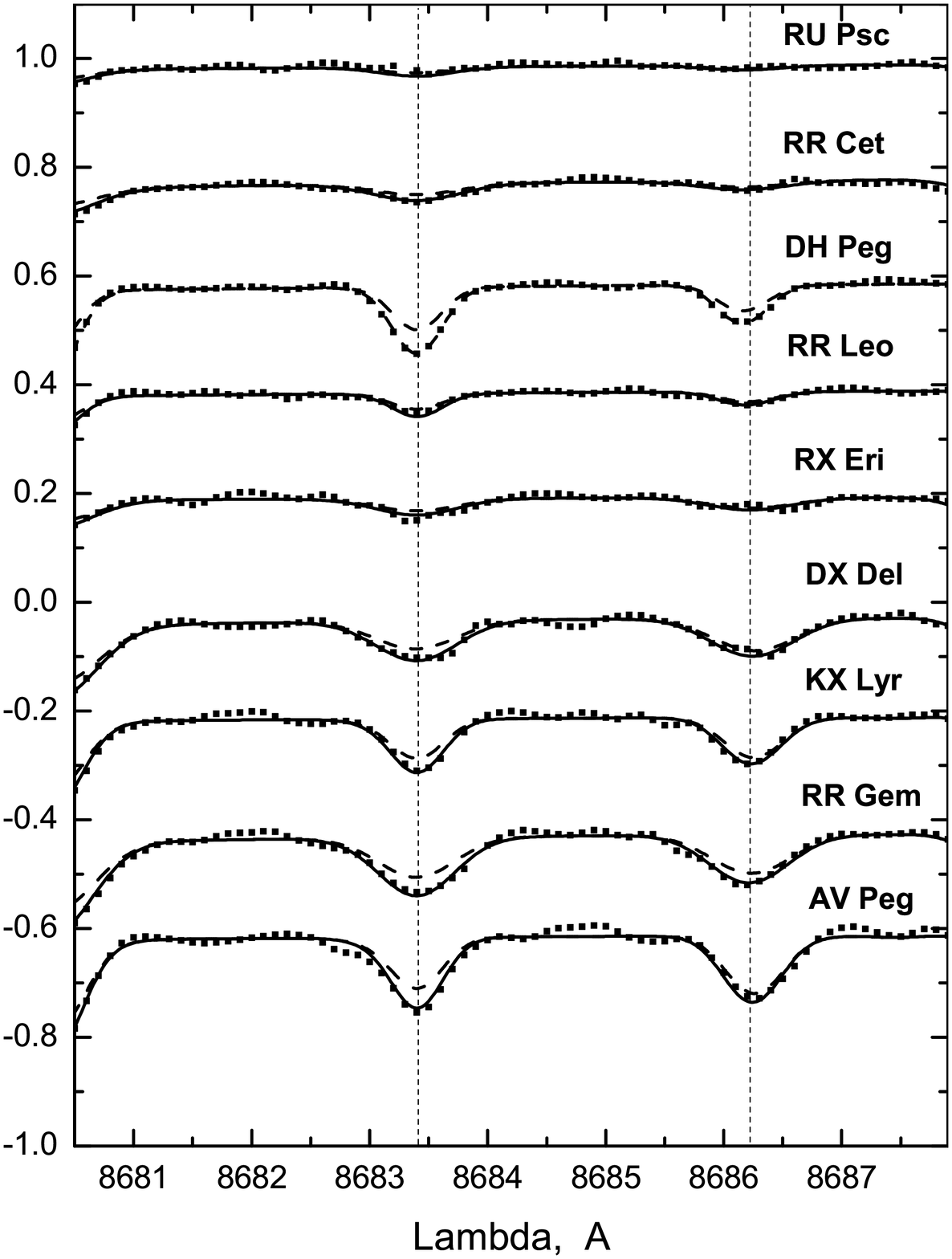}\includegraphics[width=0.66\columnwidth]{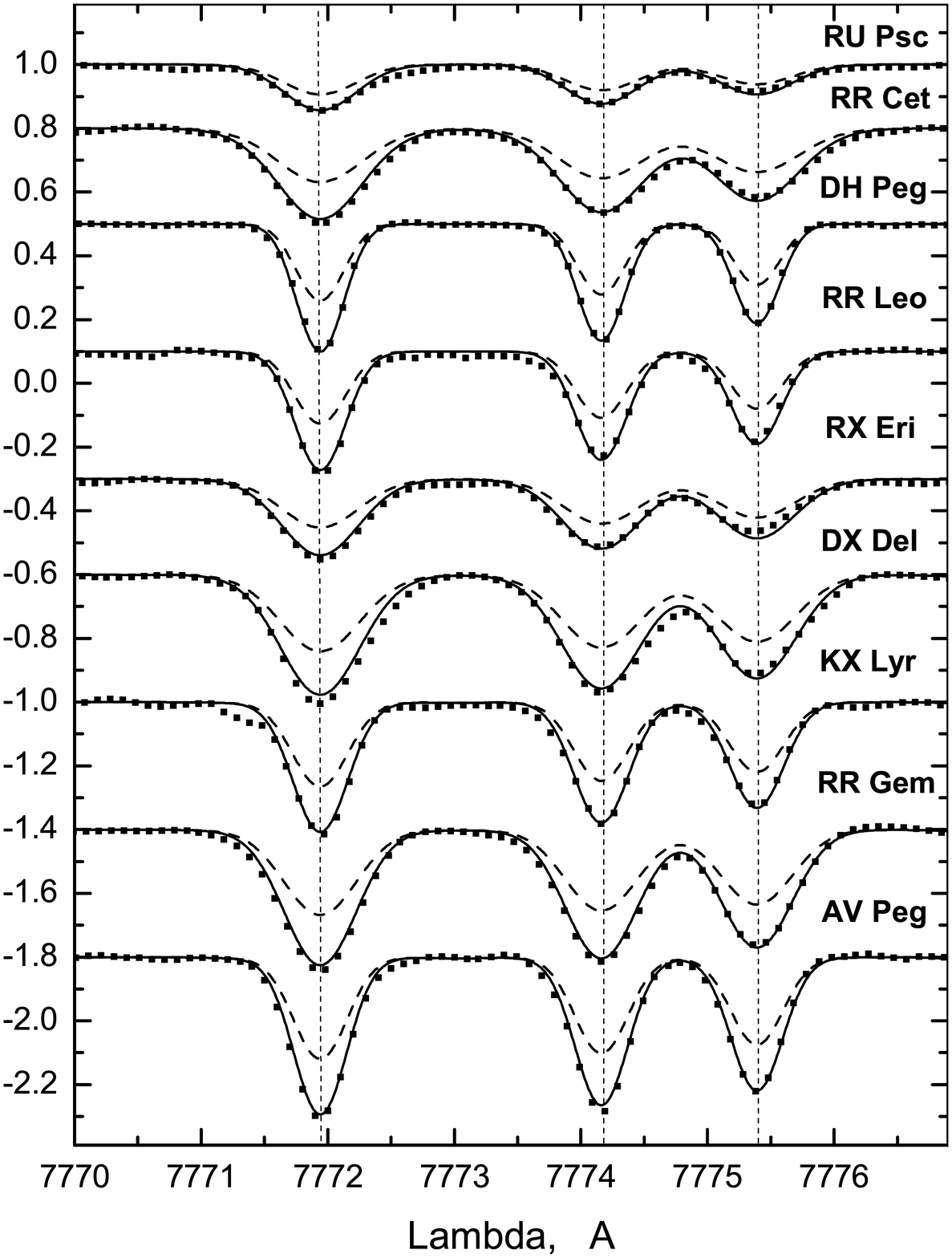}
\caption{This shows the observed C~{\sc i} lines (at 7113.18, 7115.17 and 7116.99~\AA),
N~{\sc i} lines (at  8683.40, 8686.15~\AA) and  O~{\sc i} triplet (at 7772.0,
7774.2, and 7775.4~\AA), comparing the synthesized profiles in LTE (dashed line) and NLTE (solid line) 
computed with the same abundance.}
\label{CNOlines}
\end{figure*}

\begin{figure}[t]
\centering
\includegraphics[width=0.66\columnwidth]{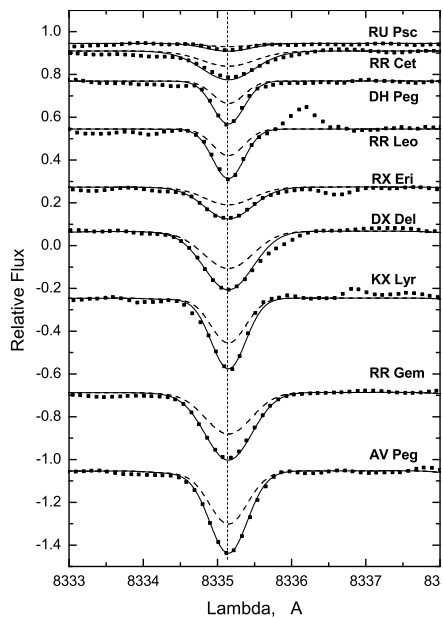}
\caption{This shows the observed C~{\sc i} lines at 8335~\AA.}
\label{C8335}
\end{figure}

\begin{figure}[t]
\centering
\includegraphics[width=0.66\columnwidth]{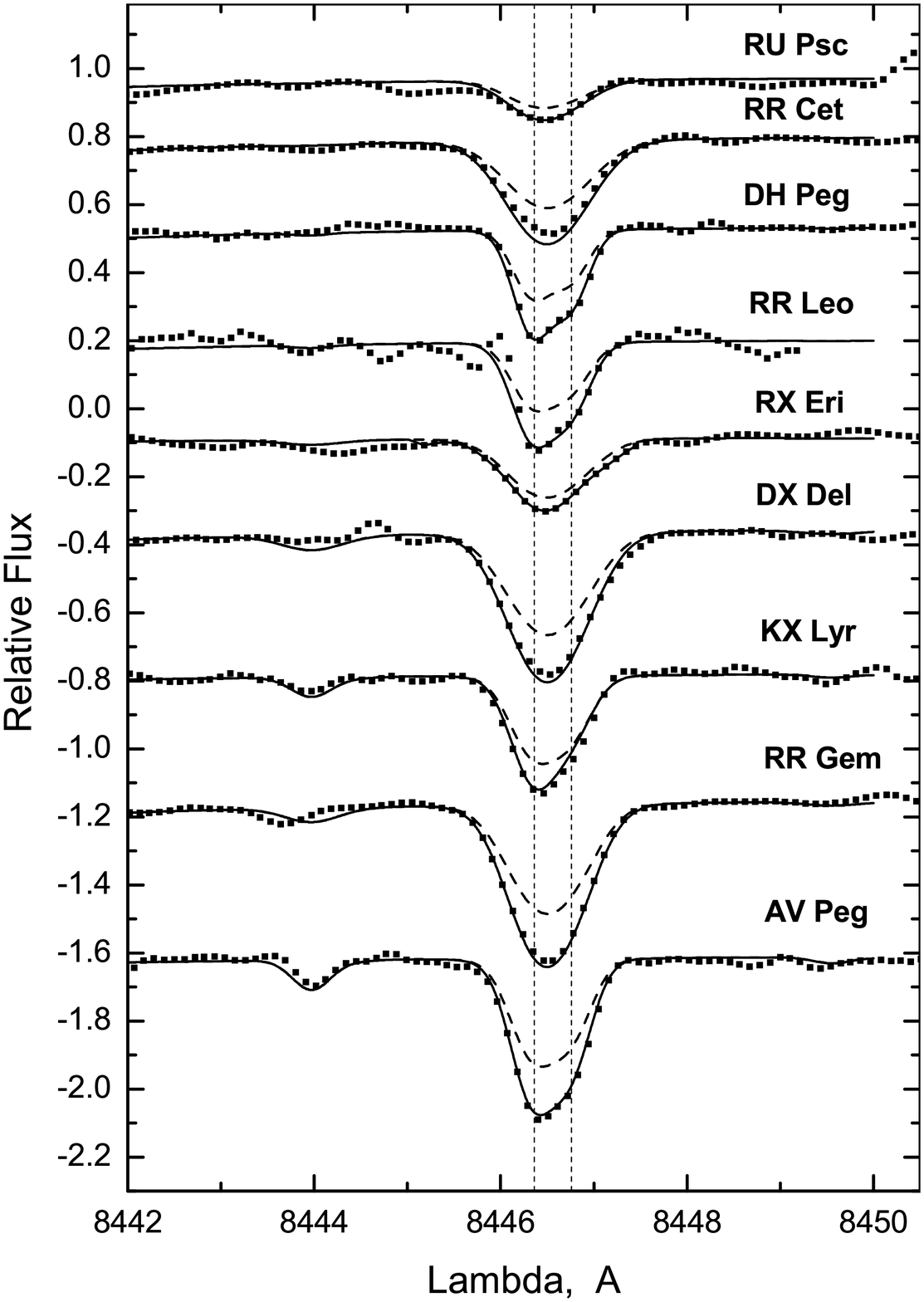}
\caption{This shows the observed O~{\sc i} lines at 8446~\AA.}
\label{O8446}
\end{figure}

In Fig. \ref{CNOlines}, Fig. \ref{C8335} and \ref{O8446} we show observed CNO lines in our program spectra together
with synthesized LTE and NLTE profiles. For RU Psc,  only the upper limit of the nitrogen abundance can be determined.
It should be noted that an error of +/-0.1 dex is inevitable in the abundance derivation 
process by using SYNTHEV, because it was done based on simple eye-judgement by trial and error on the computer display.

Table \ref{CNO_NLTE} contains for each star and each analyzed C, N, O line resulting abundance and NLTE correction
(denoted in the Table as dNLTE). Abundances and NLTE corrections in this Table are given not only for 
individual lines, but also for groups of lines located within a certain spectral range. We did not 
use in our analysis the very strong lines. Such strong lines can effectively form in the upper atmosphere 
layers, where pulsations might create specific turbulent velocity profile, which cannot be modeled in  
standard calculations. Note that correction dNLTE = LTE abundance -- NLTE abundance.

\begin{table*}
\caption[]{Abundances and NLTE corrections for all studied lines in the program spectra.}
\center
\small
\addtolength{\tabcolsep}{-1pt}
\begin{tabular}{c|cc|cc|cc|cc|cc|cc|cc|cc|cc}
\hline
\multicolumn{1}{c}{}&\multicolumn{2}{c}{DH Peg}&\multicolumn{2}{c}{RU Psc}&\multicolumn{2}{c}{AV Peg}&
\multicolumn{2}{c}{RR Gem}&\multicolumn{2}{c}{KX Lyr}&\multicolumn{2}{c}{RR Leo}&\multicolumn{2}{c}{DX Del}&
\multicolumn{2}{c}{RR Cet}&\multicolumn{2}{c}{RX Eri} \\
\hline
\multicolumn{1}{c}{C I line \AA}&
\multicolumn{1}{c}{(C/H)}&\multicolumn{1}{c}{(dNLTE)}&
\multicolumn{1}{c}{(C/H)}&\multicolumn{1}{c}{(dNLTE)}&
\multicolumn{1}{c}{(C/H)}&\multicolumn{1}{c}{(dNLTE)}&
\multicolumn{1}{c}{(C/H)}&\multicolumn{1}{c}{(dNLTE)}&
\multicolumn{1}{c}{(C/H)}&\multicolumn{1}{c}{(dNLTE)}&
\multicolumn{1}{c}{(C/H)}&\multicolumn{1}{c}{(dNLTE)}&
\multicolumn{1}{c}{(C/H)}&\multicolumn{1}{c}{(dNLTE)}&
\multicolumn{1}{c}{(C/H)}&\multicolumn{1}{c}{(dNLTE)}&
\multicolumn{1}{c}{(C/H)}&\multicolumn{1}{c}{(dNLTE)} \\
\hline            
\hline
    5052 & 6.87&  0.18      &     &         &8.32 &  0.18   &8.27 &  0.22   &7.96 &  0.17  &7.10 &  0.21   &8.06 &  0.16   &     &         &      &         \\
    5380 & 6.83&  0.16      &     &         &8.35 &  0.15   &8.31 &  0.18   &7.91 &  0.15  &     &         &8.10 &  0.14   &     &         &      &         \\
    6587 & 6.90&  0.15      &     &         &8.35 &  0.11   &8.26 &  0.13   &8.01 &  0.10  &7.04 &  0.14   &8.20 &  0.10   &     &         &      &         \\
 7111-7119& 6.87&  0.16   &     &         &8.37 &0.11-0.15&8.30 &0.15-0.20&7.96 &0.10-0.1&7.00 &0.18-0.22&8.15 &0.12-0.15&     &         &      &         \\
    8335 & 6.82&  0.42      &5.95 &  0.40   &     &         &     &         &     &        &7.00 &  0.65   &     &         &6.96 &  0.58   & 7.25 &  0.63   \\
 9061-9111& 6.80&0.55-0.9&5.99 &0.52-0.83&     &         &     &         &     &        &7.07 &0.75-1.30&     &         &7.02 &0.80-1.40& 7.29 &0.60-1.08\\
    9405 & 6.85&  0.80      &5.90 &  0.81   &     &         &     &         &     &        &7.04 &  1.36   &     &         &7.00 &  1.40   & 7.31 &  1.18   \\
 9603-9658& 6.87&0.45-0.6&6.05 &0.50-0.62&     &         &     &         &     &        &7.04 &0.55-1.03&     &         &6.99 &0.53-1.00& 7.29 &0.45-0.85\\
          &     &        &     &         &     &         &     &         &     &        &     &         &     &         &     &         &      &         \\
Mean&   6.85&        &5.97 &         &8.35 &         &8.29 &         &7.96 &        &7.04 &         &8.13 &         &6.99 &         & 7.29 &         \\
         &     &        &     &         &     &         &     &         &     &        &     &         &     &         &     &         &      &         \\
\hline
\multicolumn{1}{c}{N I line \AA}&
\multicolumn{1}{c}{(N/H)}&\multicolumn{1}{c}{(dNLTE)}&
\multicolumn{1}{c}{(N/H)}&\multicolumn{1}{c}{(dNLTE)}&
\multicolumn{1}{c}{(N/H)}&\multicolumn{1}{c}{(dNLTE)}&
\multicolumn{1}{c}{(N/H)}&\multicolumn{1}{c}{(dNLTE)}&
\multicolumn{1}{c}{(N/H)}&\multicolumn{1}{c}{(dNLTE)}&
\multicolumn{1}{c}{(N/H)}&\multicolumn{1}{c}{(dNLTE)}&
\multicolumn{1}{c}{(N/H)}&\multicolumn{1}{c}{(dNLTE)}&
\multicolumn{1}{c}{(N/H)}&\multicolumn{1}{c}{(dNLTE)}&
\multicolumn{1}{c}{(N/H)}&\multicolumn{1}{c}{(dNLTE)} \\
\hline            
\hline
 7442-7468& 7.70&  0.18  &     &         &8.14 &  0.16   &7.99 &0.20-0.24&8.05 &  0.18  &7.20 &  0.18   &7.77 &  0.15   &7.14 &  0.20   & 7.70 &  0.16   \\
 8184-8223& 7.70&0.25-0.3&6.92 &0.15-0.20&8.10 &0.15-0.20&7.97 &0.24-0.30&8.03 &0.18-0.2&7.27 &0.15-0.20&7.77 &0.14-0.19&     &         &      &         \\
    8629 & 7.70&  0.26  &     &         &8.10 &  0.20   &7.97 &  0.26   &8.00 &  0.19  &7.25 &  0.19   &7.77 &  0.16   &7.14 &  0.20   &      &         \\ 
 8683-871 & 7.70&0.25-0.3&6.92 &0.16-0.20&8.15 &0.16-0.28&7.96 &0.25-0.35&8.03 &0.19-0.2&7.24 &0.18-0.20&7.77 &0.15-0.20&7.14 &0.19-0.22& 7.70 &0.15-0.17\\
         &     &        &     &         &     &         &     &         &     &        &     &         &     &         &     &         &      &         \\
Mean  & 7.70&        &6.92 &         &8.12 &         &7.97 &         &8.03 &        &7.24 &         &7.77 &         &7.14 &         & 7.70 &         \\
         &     &        &     &         &     &         &     &         &     &        &     &         &     &         &     &         &      &         \\
\hline
\multicolumn{1}{c}{O I line \AA}&
\multicolumn{1}{c}{(O/H)}&\multicolumn{1}{c}{(dNLTE)}&
\multicolumn{1}{c}{(O/H)}&\multicolumn{1}{c}{(dNLTE)}&
\multicolumn{1}{c}{(O/H)}&\multicolumn{1}{c}{(dNLTE)}&
\multicolumn{1}{c}{(O/H)}&\multicolumn{1}{c}{(dNLTE)}&
\multicolumn{1}{c}{(O/H)}&\multicolumn{1}{c}{(dNLTE)}&
\multicolumn{1}{c}{(O/H)}&\multicolumn{1}{c}{(dNLTE)}&
\multicolumn{1}{c}{(O/H)}&\multicolumn{1}{c}{(dNLTE)}&
\multicolumn{1}{c}{(O/H)}&\multicolumn{1}{c}{(dNLTE)}&
\multicolumn{1}{c}{(O/H)}&\multicolumn{1}{c}{(dNLTE)} \\
\hline
\hline
 6156-6158& 7.99&  0.07  &     &         &8.91 &  0.06   &8.85 &  0.11   &8.67 &  0.05  &8.18 &  0.04   &8.63 &  0.07   &     &         &      &         \\
    6300 &     &        &     &         &8.92 &  0.00   &     &         &     &        &     &         &     &         &     &         &      &         \\
 7771-7775& 7.99&0.80-1.0&7.46 &0.41-0.50&8.94 &0.84-0.92&9.10 &1.07-1.15&8.67 &0.75-0.8&8.15 &0.70-0.85&8.89 &0.91-0.98&8.24 &1.00-1.24& 8.54 &0.70-0.90\\
   8446 & 7.99&  0.52  &7.46 &  0.28   &8.95 &  0.63   &9.05 &  0.73   &8.67 &  0.50  &8.13 &  0.53   &8.87 &  0.63   &8.20 &  0.70   & 8.60 &  0.52   \\
         &     &        &     &         &     &         &     &         &     &        &     &         &     &         &     &         &      &         \\
Mean  & 7.99&        &7.46 &         &8.93 &         &9.00 &         &8.67 &        &8.15 &         &8.80 &         &8.22 &         & 8.57 &         \\                                                         
\hline 
\end{tabular}
\label{CNO_NLTE}
\end{table*}                                                                                                                                            

In Table \ref{CNO_NLTE} we give NLTE corrections as a specific value, when in the indicated range there is only one line. If there are several lines,
we show some range of corrections (say, 0.55 -- 0.90). As a rule such lines belong to the same multiplet with different intensities. The entire ensemble of this lines
can be described with a single NLTE abundance. However, these lines have different NLTE corrections. It is the range of corrections for different 
lines from the certain spectral range, and it is shown in this Table. For instance, among the lines of IR oxygen triplet for the star DH Peg, the strongest line
has NLTE correction  --1.05, while the weakest line has NLTE correction --0.80. Thus, in the Table we indicated the range 0.80--1.05 for the spectral 
range 7771--7775 \AA.

\begin{figure}[t]
\centering
\includegraphics[width=1.0\columnwidth]{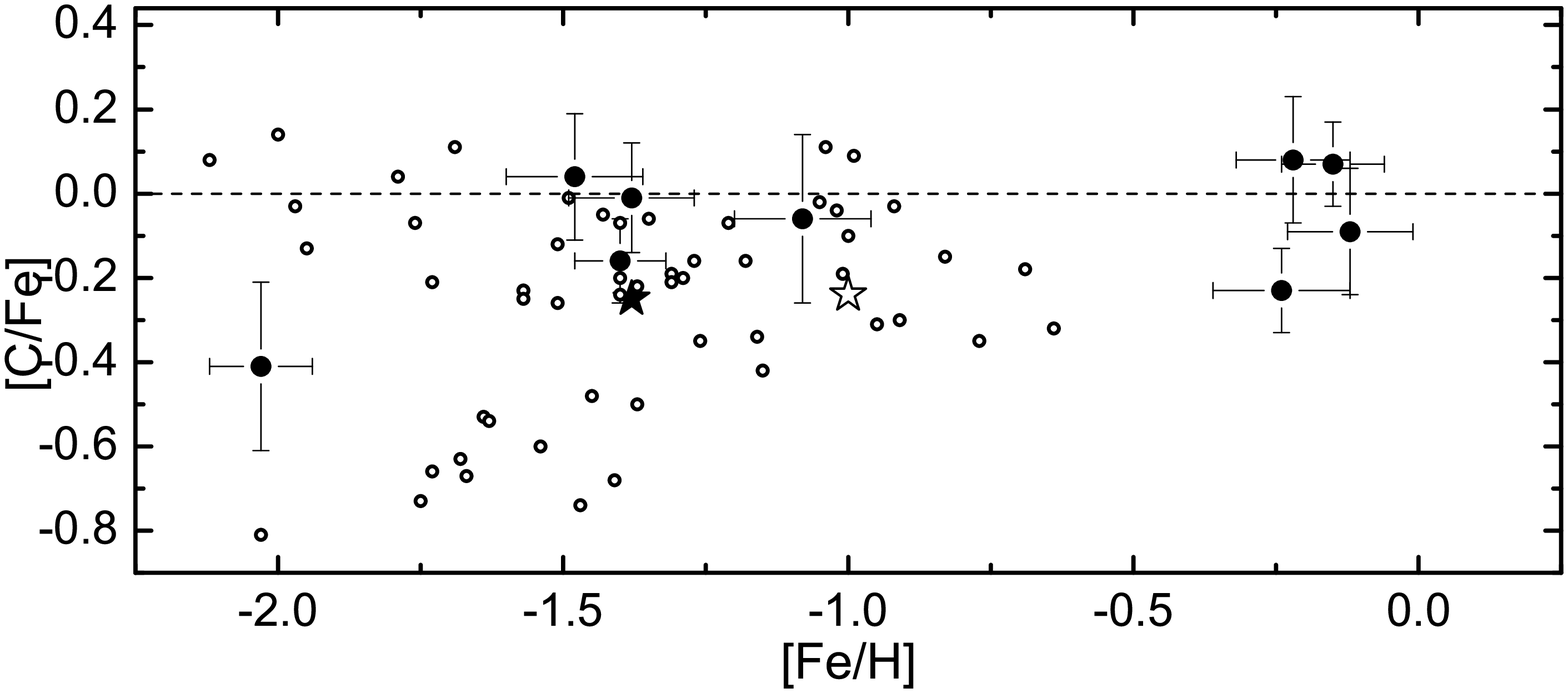}
\includegraphics[width=1.0\columnwidth]{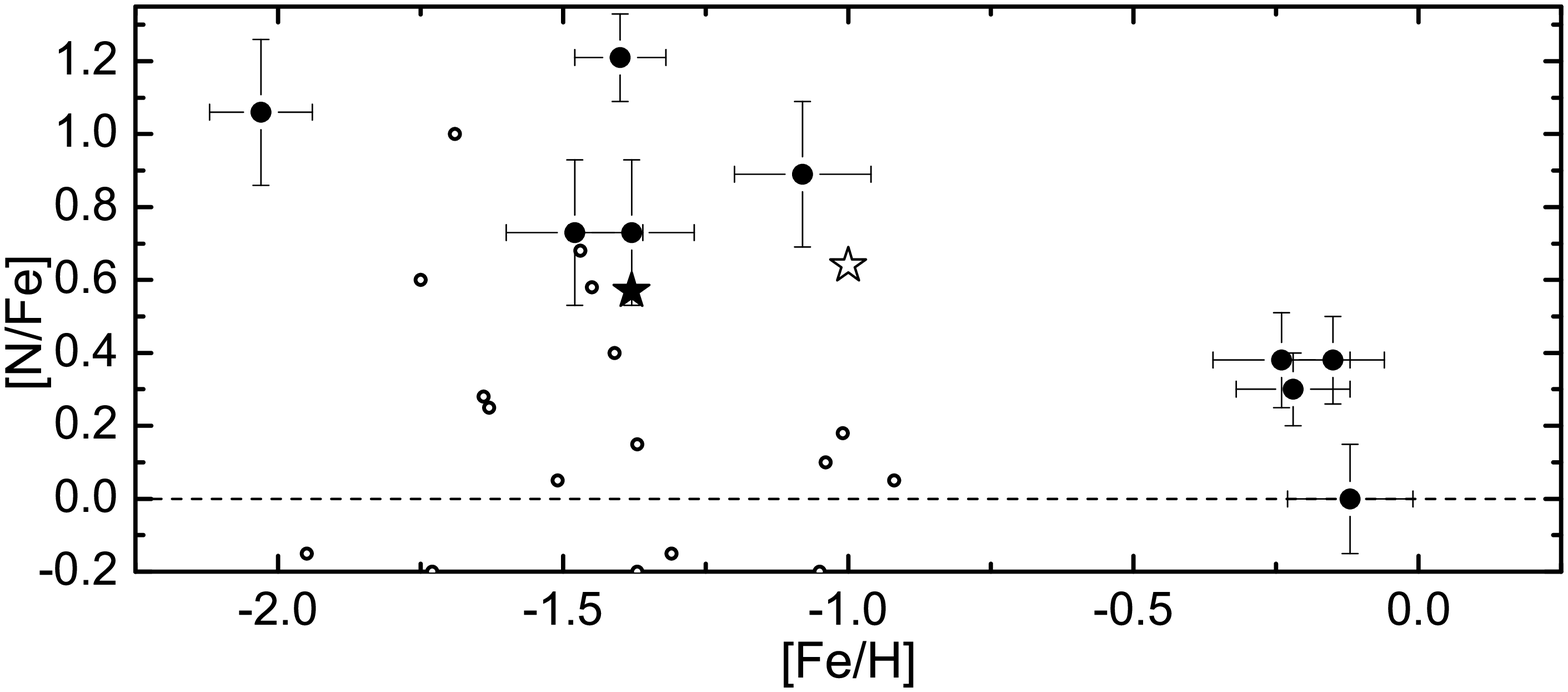}
\includegraphics[width=1.0\columnwidth]{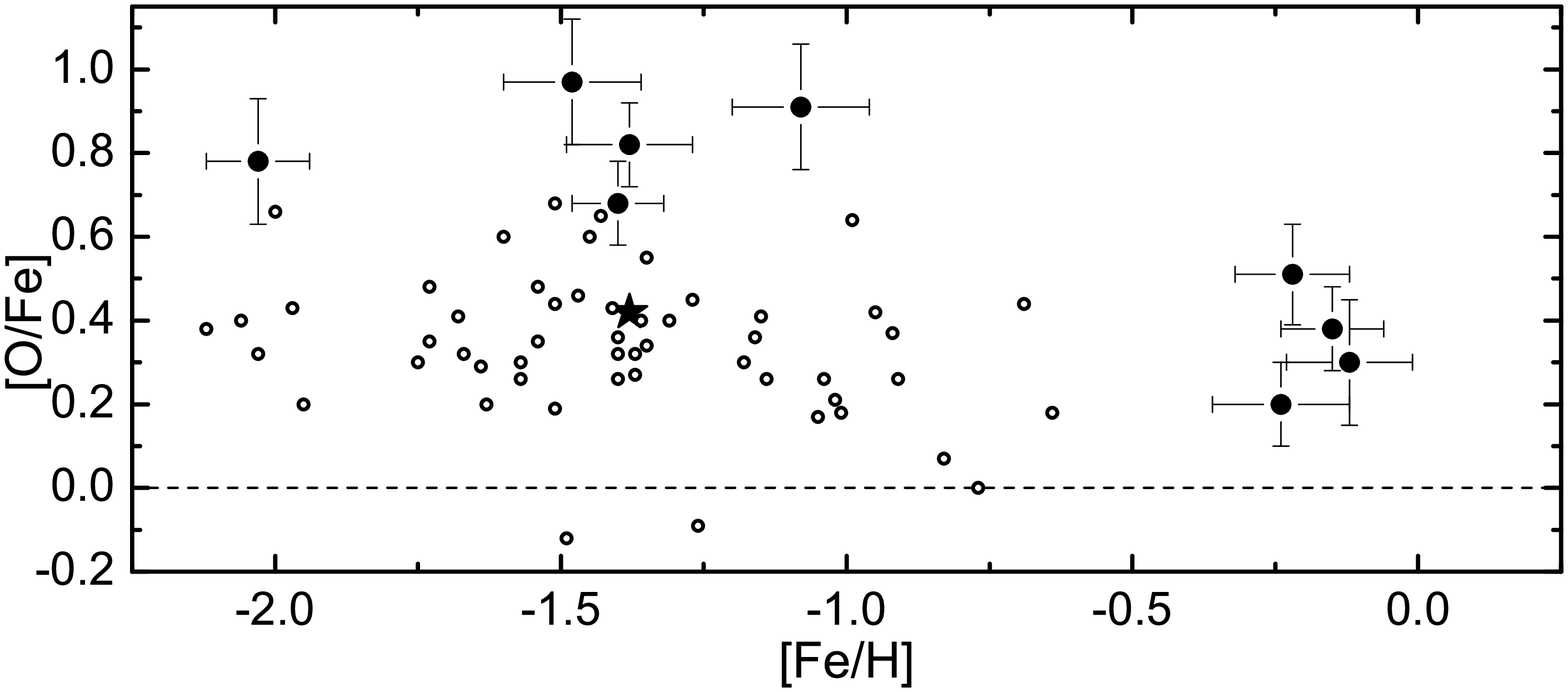}
\caption{Relative carbon, nitrogen and oxygen abundance vs. metallicity in 
our program stars.The star No. 4408 from M4 is designated as an open asterisk symbol, 
HD~161817 is shown as a solid asterisk symbol. Small open circles designate the data
from \citet{Gratton2000}. Error bars for our data are indicated. The error bar includes uncertainty in the atmosphere parameter determination 
(typical errors for carbon and oxygen were discusseed in \citealt{Andrievsky2020}, and reach about 0.08 dex and 0.12 dex respectively). Similar 
uncertainty for nitrogen lines is about 0.12 dex. In addition, the error bar contains uncertainty of the profile fit (independent error). 
The resulting error is the root of the sum squares.} 

\label{CNO-Fe}
\end{figure}

\section{Results and Discussion}

In our previous paper \citep{Andrievsky2020} we reported about decreased/or
near-to-normal carbon abundance in a sample of 21 field RR Lyrae stars.
We made a conclusion that no addtional carbon is expected to be brought to the 
stellar surface after the He flash in the RR Lyrae core. Nitrogen abundance
was not derived for those stars.
 
From the top panel of Fig. \ref{CNO-Fe}we see that our current NLTE result 
confirms our previous conclusion. Middle panel of Fig. \ref{CNO-Fe} shows 
that NLTE nitrogen abundance is clearly increased in RR Lyrae stars. 
It is natural to assume that on the surface of the studied stars we see 
a sign of the material, which was processed in an incomplete CNO-cycle. 
The bottom panel shows the behavior of the oxygen abundance in RR Lyrae stars, 
qualitatively typical for the Population II stars. However,
two stars with close-to-solar metallicity, AV Peg and RR Gem, show significantly 
increased oxygen abundance (+0.38 and +0.51 respectively). If we derived
the wrong atmosphere parameters for these stars, our metallicity ([Fe/H]) should
obviously be incorrect, since parameters can change significantly during the pulsation cycle, 
while metallicity should be the same within the standard error for each pulsation
phase (see, for instance, \citealt{For2011,Liu2013}). However, we see that our 
metallicity estimate for AV Peg is almost the same as that obtained by \citet{Chadid2017}. 
These authors report for AV Peg [Fe/H] = --0.14, while we obtained [Fe/H] = --0.15. 

For RR Gem the difference with literature data on [Fe/H] is larger, but
still within acceptable limits. Thus, in the paper of \citet{Magurno2018} 
authors give the value [Fe/H] = --0.41, while our value is 0.19 dex higher. 
Of course, many factors may affect the abundance value, but such 
a difference is not critical and cannot significantly reduce the final oxygen 
abundance.

Together with our program stars we show in Fig. \ref{CNO-Fe} the stars,
which have been analyzed by  \citet{Gratton2000}. These authors used
molecular bands to derive the C and N abundances, as well as forbidden and permitted 
lines for oxygen. The spectra of stars located at different evolutionary
stages were used: main sequence, lower and upper red giant branch and
red horizontal branch. 

Of particular interest for our study is to compare the abundances of C and N in 
the stars of our program with the abundances of these elements in non-variable 
HB stars. Following \cite{Behr2005} we can note the various
regions in the horizontal branch. The red part of the HB (RHB) with
temperature boundaries from about 4500 to 6000 K.
At the higher temperatures there is a domain of RR Lyrae type stars.
Their instability strip comprises temperature range from 6000 to 7500 K.
Toward the higher temperatures, the cool end of the hot blue HB (BHB)
is located. It covers effective temperature range from
7500 to 11500 K. Finally, the hot part of the BHB stretches
from 11500 K to the higher temperatures.

There are works devoted to the light elements abundance study in the RHB stars.
Among them, for instance, \cite{Gratton2000},
who studied ten RHB stars bordering the group of stars in the 
lower red giant branch. The authors reported about moderately increased
nitrogen and decreased carbon abundance in their sample of stars.
A similar conclusion was also reached by \cite{Tautvaisiene2001}
who studied intermediate metal deficient cool core helium burning stars from
the Galactic thick disc. Moreover, the authors of that paper gave additional
evidence that mixing processes in the giant stars can be metallicity dependent.
Both studies used CH and CN molecular bands to determine abundances of carbon
and nitrogen. At temperatures of the RHB stars atomic
C and N lines are not visible in their spectra. Since our NLTE analysis
of RR Lyrae stars was based on the use of atomic lines, we tried fo
find the literature spectroscopic information for the
stars hotter than RR Lyrae variables. Since, as is known \citep{Michaud1983},
in the stars hotter than 11500 K atomic diffusion processes
can change the surface abundances of elements including carbon and nitrogen,
we focused on finding the appropriate spectroscopic data for the stars
lying in the temperature range from 7500 to 11500 K. Unfortunately, this 
temperature range, as a rule, interested specialists mainly because of the 
helium lines, for which only small fragments of blue spectra were observed.

The necessary data we found in the work of \cite{Lambert1992}.
They investigated three stars, and for one of them (HB stars No. 4408 from M4)
one high-resolution spectum in the near-infrared range (from 8080 to 9620 \AA)
was exposed using CTIO facilities.
Fortunately, authors give in their paper equivalent widths of carbon and
nitrogen lines. Using the NLTE approximation and published equivalent widths,
we esimated absolute carbon abundance for this star (C/H) = $7.19 \pm 0.06$ 
that corresponds to [C/Fe] = --0.24. The result for nitrogen is as follows:
(N/H) = $7.53 \pm 0.10$,  [N/Fe]=+0.64.
It should be noted that average NLTE corrections for investigated lines are
significant:
--040 for carbon, and --0.55 for nitrogen. Note that we used iron abundance
for this cluster [Fe/H] = --1.0, as reported by \cite{Lambert1992}.

Another good example for us of a star located in HB to the left of the instability 
strip is HD 161817. \cite{Takeda1997}, hereinafter T\&S, calculated the CNO 
abundances in this star and the corresponding NLTE corrections using the NLTE 
approximation. The authors used two atmosphere models: \Teff = 7500 K, \logg=3.0,
one with [Fe/H]=--1 and another with [Fe/H]=--2. The authors derived resulting 
metallicity as: (Fe/H)= 6.12, i.e. [Fe/H] = --1.38, and the following abundances: 
(C/Fe)=6.9, (N/Fe)=7.0 and (O/Fe)=7.8.  We used calculated model atmosphere 
for the avove parameters and metallicity,  [Fe/H] = --1.38. With our NLTE code, 
atomic models and published equivalent widths from T\&S we obtained: 
(C/Fe)=6.80, (N/Fe)=7.08 and (O/Fe)=7.75. These values are close to the
T\&S values.

\begin{table}
\caption[]{NLTE corrections for CNO abundances.}
\center
\begin{tabular}{ccc}
\hline
C~{\sc i} \\
\hline
Line, \AA &           T\&S    &  Present work \\
\hline
9078      & --0.31, --0.61 &     --0.55       \\
9088      & --0.32, --0.62 &     --0.55       \\
9095      & --0.81, --1.23 &     --1.05       \\            
\hline
N~{\sc i} \\
\hline
8683      & --0.42, --0.71 &     --0.25       \\
8686      & --0.37, --0.65 &     --0.22       \\
8706      & --0.33, --0.59 &     --0.24       \\
\hline
O~{\sc i} \\
\hline
IR triplet & --0.65, --0.78 &     --0.70       \\
\hline
\end{tabular}
\\

Remark: in T\&S corrections for two models are given.
\label{TS}
\end{table}

As one can see from Table \ref{TS}, our corrections and corrections of T\&S 
are close enouugh, except for nitrogen. We investigated the reason of 
this discrepancy. First, like T\&S, we calculated NLTE and LTE nitrogen abundances 
for all the lines using  two atmosphere models (one for [Fe/H] = --1, and 
the other for [Fe/H] = --2). Then we compared our results with those of T\&S.
All results are given in Table \ref{TS_N}.

\begin{table*}
\caption[]{Toward a solution to the problem of the NLTE nitrogen corrections.}
\center
\begin{tabular}{c|cccccc|cccccccccccc}
\hline
\multicolumn{1}{c}{}&\multicolumn{6}{c}{T\&S}&\multicolumn{5}{c}{Our data}\\
\hline
\multicolumn{1}{c}{Line, \AA}&\multicolumn{2}{c}{NLTE}&\multicolumn{2}{c}{LTE}
&\multicolumn{2}{c}{corr}&\multicolumn{1}{c}{NLTE}&\multicolumn{1}{c}{LTE}
&\multicolumn{1}{c}{corr}&\multicolumn{2}{c}{LTE}\\
\hline
\multicolumn{1}{c}{}&\multicolumn{1}{c}{mod1}&\multicolumn{1}{c}{mod2}
&\multicolumn{1}{c}{mod1}&\multicolumn{1}{c}{mod2}&\multicolumn{1}{c}{mod1}
&\multicolumn{1}{c}{mod2}
&\multicolumn{3}{c}{our model}
&\multicolumn{1}{c}{mod1}&\multicolumn{1}{c}{mod2}\\
\hline
8683   & 7.06 & 6.78 & 7.48 & 7.49 & --0.42 & --0.71 & 6.92 & 7.17 & --0.25 & 7.14 & 7.17 \\
8686   & 7.16 & 6.89 & 7.53 & 7.54 & --0.37 & --0.65 & 7.07 & 7.29 & --0.22 & 7.25 & 7.27 \\
8706   & 7.28 & 7.03 & 7.61 & 7.62 & --0.33 & --0.59 & 7.25 & 7.49 & --0.24 & 7.46 & 7.47 \\
\hline
\label{TS_N}
\end{tabular}

Remark: our NLTE and LTE abundances (columns 8 and 9) were derived using a single model
with [Fe/H] = --1.38. In addtion, we nobtained LTE abundances using two models, similar to
T\&S. The results are given in columns 11 and 12.

\end{table*}

By comparing the NLTE and LTE data from Table \ref{TS_N}, we can conclude that we agree
with T\&S on the NLTE results, while their LTE data (and NLTE corrections, respectively) 
seem to be incorrect for an unknown reason. Since our LTE data were derived using the 
MULTI code, we decided to check ourselves and applied Kurucz's WIDTH6 code with models 
of 1993 and equivalent widths from T\&S. We got the folloowing result: WIDTH6 and MULTI LTE 
abundances agree within 0.03 dex.

Finally, it should be noted that T\&S used solar CNO abundances
(C/H)=8.6, (N/H)=8.0, (O/H)=8.9, which is slightly different from 
the currently adopted values: 8.43, 7.89 and 8.71 respectively.

C,N (and O) abundance data for two hot HB stars are plotted in Fig. \ref{CNO-Fe}.

\begin{figure}[t]
\centering
\includegraphics[width=1.0\columnwidth]{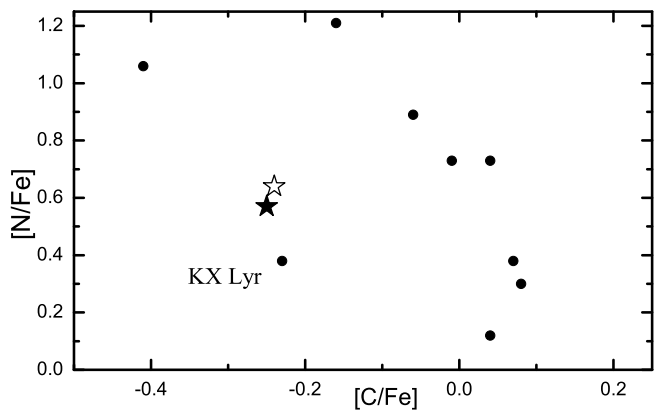}
\caption{Relative nitrogen abundance [N/Fe] as a function of [C/Fe].
The star No. 4408 from M4 is designated as an open asterisk symbol, 
HD~161817 is shown as a solid asterisk symbol.
One deviating point KY Lyr is indicated.}
\label{C-N}
\end{figure}

In Fig. \ref{C-N} we show the distribution of [N/Fe] versus [C/Fe]
for all program stars including M4 HB star  and HD~161817.
There is a fairly clear increase in the relative nitrogen abundance with
decreasing carbon abundance. This figure allows us to conclude that 
we see at the surface of the RR Lyrae stars the results of the previous 
first dredge up episode that occured at the RGB, when the material from 
the incomplete CNO cycle was brought to the upper atmosphere. 

\begin{figure}[t]
\centering
\includegraphics[width=1.0\columnwidth]{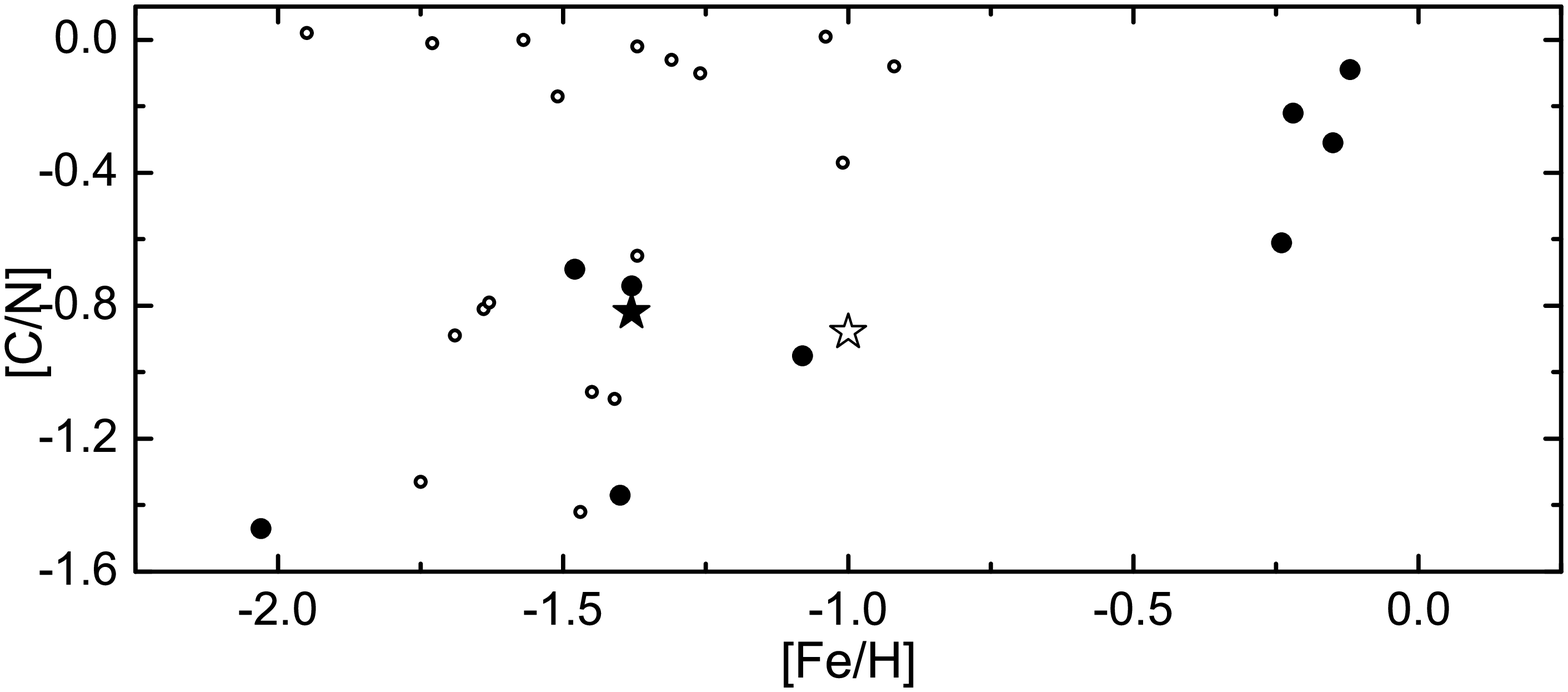}
\caption{The ratio [C/N] as a function of metallicity.
The designations are the same as in Fig. \ref{CNO-Fe}.}
\label{N-Fe3}
\end{figure}

In Fig. \ref{N-Fe3}we show the ratio [C/N] as a function
of metallicity. Our data clearly demonstrate a gradual decrease of the
relative abundance of C/N as metallicity decreases in a wide range. 

Let us briefly discuss Fig. \ref{N-Fe3}. \cite{Lagarde2019} examined the effect
of thermohaline mixing on observed surface abundances of carbon and nitrogen
(thermohaline instability triggering an effective additional mixing in giant stars
was first considered by \citealt{Charbonnel2007}). The authors used a stellar population 
synthesis model and compared their synthetic data with the observed [C/N] ratios available 
through abundance determinations from the Gaia-ESO survey. The authors showed that 
thermohaline mixing is an efficient mechanism that can alter surface abundances 
of carbon and nitrogen, especially in stars with lower metallicity. Their Fig. 8 shows 
that theoretical prediction of the [C/N] behaviour indicates that this ratio should decrease 
as metallicity decreases. Partially, this prediction is supported by observations of the 
cluster stars. For instance, [C/N] data for old globular cluster NGC~1851 and open cluster 
M~67 do not contadict the above prediction. 

Our sample of stars covers the metallicity range from about --0.2 to --2 dex. 
Corrspondingly, [C/N] ratio ranges from about zero to --1.5 dex, which is in
good agreement with \cite{Lagarde2019} results from their stellar population synthesis
(see their Fig. 8).

\section{Conclusion}

We have derived the NLTE abundances of carbon, nitrogen and oxygen in a
sample of RR Lyrae stars. The iron content in LTE in these stars ranges
from --0.15 to --2.03. The distribution of oxygen abundance in this
sample is typical of halo stars: the lower the metallicity, the higher
the oxygen abundance. The carbon in our program stars is not increased.
Both of these findings are similar to those reported by \cite{Andrievsky2020}
for a larger sample of RR Lyrae stars. 

Nitrogen is clearly overabundant in the studied RR Lyrae stars; its abundance
increases with decreasing carbon abundance. Thus, we conclude that our stars 
exhibit material on their surface that was processed in an incomplete CNO cycle, 
when the stars passed the RGB stage, and then mixed due to convection with 
atmospheric gas during the first dredge up. 

The ratio [C/N] distribution versus [Fe/H] for our program RR Lyrae type stars 
is similar to that observed for halo population giants, where the surface C and N anundances
are effectively altered by additional mixing induced by thermohaline instability.
Our result on RR Lyrae stars shows that this process is active, and confirms the 
conclusion that thermohaline mixing becomes more effective in the giant stars with 
lower metallicity \citep{Lagarde2011}.


\section*{Acknowledgments}

We are grateful to Prof. George Wallerstein for inspiring us at one time to
carry out spectroscopuic researches on RR Lyrae stars, and he himself took
an active part in this process. Special thanks to W.~Huang for his help with
a preliminary reduction of the program spectra.
We are grateful to our referee for the numerous valuable comments, which we
hope have improved the content of this paper.

\bibliography{RR_Lyr_CNO_AN}%

\newpage

\appendix




\section{On-line material}


\begin{table*}
\center
\caption[]{Equivalent widths of the used iron lines and corresponding abundances for DH Peg, RU Psc, AV Peg, 
RR Gem, KX Lyr. Atomic data are from VALD.}
\begin{tabular}{rlrrrrrrr}
\hline
 Lambda   &    Ion      &  EPL  & \loggf  &       DH Peg  &         RU Psc   &        AV Peg    &       RR Gem   &      KX Lyr   \\
\hline
\hline
 4602.001 & Fe {\sc i}  & 1.608 & --3.153 &   -- \ \ \ \ \ --  &     -- \ \ \ \ \ --   &     -- \ \ \ \ \ --   &   31    7.22   &    -- \ \ \ \ \ -- \\
 4602.941 & Fe {\sc i}  & 1.485 & --2.208 &   -- \ \ \ \ \ --  &     10    5.25   &     -- \ \ \ \ \ --   &   -- \ \ \ \ \ --   &    -- \ \ \ \ \ -- \\
 4620.521 & Fe {\sc ii} & 2.828 & --3.190 &   -- \ \ \ \ \ --  &     16    5.37   &     -- \ \ \ \ \ --   &   -- \ \ \ \ \ --   &    -- \ \ \ \ \ -- \\
 4625.045 & Fe {\sc i}  & 3.241 & --1.270 &   -- \ \ \ \ \ --  &     -- \ \ \ \ \ --   &     -- \ \ \ \ \ --   &   75    7.39   &    -- \ \ \ \ \ -- \\
 4690.138 & Fe {\sc i}  & 3.686 & --1.680 &   -- \ \ \ \ \ --  &     -- \ \ \ \ \ --   &     -- \ \ \ \ \ --   &   24    7.33   &    -- \ \ \ \ \ -- \\
 4731.453 & Fe {\sc ii} & 2.891 & --3.100 &   -- \ \ \ \ \ --  &     16    5.33   &     -- \ \ \ \ \ --   &   -- \ \ \ \ \ --   &    -- \ \ \ \ \ -- \\
 4733.592 & Fe {\sc i}  & 1.485 & --2.987 &   -- \ \ \ \ \ --  &     -- \ \ \ \ \ --   &     -- \ \ \ \ \ --   &   38    7.06   &    -- \ \ \ \ \ -- \\
 4736.773 & Fe {\sc i}  & 3.211 &  --.670 &   -- \ \ \ \ \ --  &     18    5.53   &     -- \ \ \ \ \ --   &  106    7.31   &    -- \ \ \ \ \ -- \\
 4741.530 & Fe {\sc i}  & 2.832 & --2.000 &   -- \ \ \ \ \ --  &     -- \ \ \ \ \ --   &     -- \ \ \ \ \ --   &   38    7.20   &    -- \ \ \ \ \ -- \\
 4745.800 & Fe {\sc i}  & 3.654 & --1.269 &   -- \ \ \ \ \ --  &     -- \ \ \ \ \ --   &     -- \ \ \ \ \ --   &   42    7.23   &    -- \ \ \ \ \ -- \\
 4788.757 & Fe {\sc i}  & 3.237 & --1.810 &   -- \ \ \ \ \ --  &     -- \ \ \ \ \ --   &     -- \ \ \ \ \ --   &   35    7.30   &    -- \ \ \ \ \ -- \\
 4892.859 & Fe {\sc i}  & 4.218 & --1.289 &   -- \ \ \ \ \ --  &     -- \ \ \ \ \ --   &     40    7.50   &   20    7.26   &    28    7.28 \\
 4893.820 & Fe {\sc ii} & 2.828 & --4.266 &    8    6.17  &     -- \ \ \ \ \ --   &     58    7.41   &   43    7.18   &    47    7.31 \\
 4917.230 & Fe {\sc i}  & 4.191 & --1.179 &   -- \ \ \ \ \ --  &     -- \ \ \ \ \ --   &     53    7.57   &   34    7.43   &    39    7.35 \\
 4923.927 & Fe {\sc ii} & 2.891 & --1.319 &   -- \ \ \ \ \ --  &     -- \ \ \ \ \ --   &     -- \ \ \ \ \ --   &   -- \ \ \ \ \ --   &    -- \ \ \ \ \ -- \\
 4924.770 & Fe {\sc i}  & 2.279 & --2.240 &    9    6.13  &     -- \ \ \ \ \ --   &     -- \ \ \ \ \ --   &   60    7.31   &    81    7.37 \\
\hline
\end{tabular}
\label{spectra}
\end{table*}


\begin{table*}
\center
\caption{Atomic data, equivalent widths of iron lines and corresponding abundances for  RR Leo,
DX Del, RR Cet,  RX Eri.}
\begin{tabular}{rlrrrrrrr}
\hline
\hline
Lambda    &    Ion      &  EPL   &  \loggf    &   RR Leo      &   DX Del        &  RR Cet       &    RX Eri     \\
\hline
 4598.117 & Fe {\sc i}  &  3.283 & --1.569    &  -- \ \ \ \ \ --  &    -- \ \ \ \ \ --  &   -- \ \ \ \ \ --   &    26    6.53 \\
 4602.941 & Fe {\sc i}  &  1.485 & --2.208    &  -- \ \ \ \ \ --  &    -- \ \ \ \ \ --  &   -- \ \ \ \ \ --   &    85    6.55 \\
 4619.288 & Fe {\sc i}  &  3.603 & --1.030    &  -- \ \ \ \ \ --  &    -- \ \ \ \ \ --  &   -- \ \ \ \ \ --   &    31    6.39 \\
 4620.521 & Fe {\sc ii} &  2.828 & --3.190    &  -- \ \ \ \ \ --  &    -- \ \ \ \ \ --  &   -- \ \ \ \ \ --   &    59    6.35 \\
 4625.045 & Fe {\sc i}  &  3.241 & --1.270    &  -- \ \ \ \ \ --  &    -- \ \ \ \ \ --  &   -- \ \ \ \ \ --   &    28    6.25 \\
 4728.546 & Fe {\sc i}  &  3.654 & --1.280    &  -- \ \ \ \ \ --  &    -- \ \ \ \ \ --  &   -- \ \ \ \ \ --   &    25    6.55 \\
 4733.592 & Fe {\sc i}  &  1.485 & --2.987    &  -- \ \ \ \ \ --  &    -- \ \ \ \ \ --  &   -- \ \ \ \ \ --   &    27    6.34 \\
 4736.773 & Fe {\sc i}  &  3.211 & --0.670    &  -- \ \ \ \ \ --  &    -- \ \ \ \ \ --  &   -- \ \ \ \ \ --   &    75    6.41 \\
 4741.530 & Fe {\sc i}  &  2.832 & --2.000    &  -- \ \ \ \ \ --  &    -- \ \ \ \ \ --  &   -- \ \ \ \ \ --   &    15    6.26 \\
 4745.800 & Fe {\sc i}  &  3.654 & --1.269    &  -- \ \ \ \ \ --  &    -- \ \ \ \ \ --  &   -- \ \ \ \ \ --   &    15    6.28 \\
 4892.859 & Fe {\sc i}  &  4.218 & --1.289    &  -- \ \ \ \ \ --  &     33        7.44  &   -- \ \ \ \ \ --   &   -- \ \ \ \ \ -- \\
 4893.820 & Fe {\sc ii} &  2.828 & --4.266    &  -- \ \ \ \ \ --  &     52        7.31  &   -- \ \ \ \ \ --   &    12    6.45 \\
 4917.230 & Fe {\sc i}  &  4.191 & --1.179    &  -- \ \ \ \ \ --  &     43        7.48  &   -- \ \ \ \ \ --   &   -- \ \ \ \ \ -- \\
 4924.770 & Fe {\sc i}  &  2.279 & --2.240    &  -- \ \ \ \ \ --  &     78        7.42  &   -- \ \ \ \ \ --   &    29    6.35 \\
 4950.106 & Fe {\sc i}  &  3.417 & --1.669    &  8    6.28        &    -- \ \ \ \ \ --  &   -- \ \ \ \ \ --   &    15    6.46 \\
\hline
\label{EWs}
\end{tabular}
\end{table*}


\label{lastpage}

\end{document}